# Effects of Pore Connectivity and Tortuosity on the Dynamics of Fluids Confined in Sub-nanometer Pores

Siddharth Gautam[*,1], David R. Cole[1]

[1]*School of Earth Sciences, The Ohio State University, 275 Mendenhall Laboratory, 125 South Oval Mall, Columbus, OH 43210 USA*

## Abstract

Dynamical behavior of fluids under nano-pore confinement is studied extensively as it has important implications for several industrial as well as geological processes. Pore network in many porous materials exhibits a varied degree of inter connections. The extent of this pore connectivity may affect the structural and dynamical behavior of the confined fluid. However, studies of fluid confinement addressing these effects systematically are lacking. Here, we report molecular dynamics simulation studies addressing the effects of pore connectivity on the dynamics of two representative fluids – $CO_2$ and ethane in silicalite by systematically varying the degree of pore connectivity through selectively blocking some pore space with immobile methane molecules. By selectively turning off the pore spaces in the shape of straight, or tortuous zigzag channels, we also probe the effects of pore tortuosity. In general, pore connectivity is found to facilitate both the translational as well as rotational dynamics of both fluids, while the intermolecular modes of vibration in both fluids remain largely unaffected. The effects of providing connections between a set of straight or zigzag channel-like pores are however more nuanced. Pore tortuosity facilitates the rotational motion, but suppresses the translational motion of $CO_2$, while its effects on the rotational and translational motion of ethane are less pronounced. The intermolecular vibrational modes of both fluids shift to higher energies with an increase in the number of tortuous pores. The results reported here provide a detailed molecular level understanding of the effects of pore connectivity on the dynamics of fluids and thus have implications for applications like fluid separation.

*Corresponding author, email: gautam.25@osu.edu

## 1.0 INTRODUCTION

Migration of fluids through the lithosphere occurs through porous rocks of varying porosity and permeability [1]. A significant fraction of these pores can occur at dimensions less than 100 nm and contribute significantly to accessibly reactive surface area [2]. When fluids transport via these pores, their dynamical behavior deviates significantly from the behavior in bulk. For this reason, dynamical behavior of fluids confined in nanopores has been an important field of investigation [1,3]. The scenario of fluids confined in nano-scale pores also plays out in various industrial applications including catalysis [4]. Indeed, confining reactants in nano-pores has been a useful catalysis strategy [5]. Storage of $CO_2$ and other gases is another application where understanding the behavior of fluids under nano-confinement is important [6]. Membrane-based separation of fluids also makes use of the fact that some fluids can pass through the nanopores in a membrane more efficiently than others [7].

Pores in natural as well as engineered materials can exhibit a varied degree of inter-connectivity [8]. Some engineered porous media may exhibit unidimensional pores isolated from each other. MCM-41 and ZSM-22 are examples of such porous media with unidimensional pores [9, 10]. Fluid dynamics in such unidimensional pores is predominantly unrestricted in the direction of pore axis and is hindered mainly by the presence of another species. For example, propane is found to be hindered by the presence of water bridges in MCM-41-S pores [11, 12]. On the other hand, the pore structure in natural media is more disordered and can have pore spaces that are blocked and isolated from other pores resulting in a variation in the degree of inter-connectivity in pores [13]. This variation in the degree of inter-connectivity of pore spaces may affect the behavior of fluids confined in them [14 – 18]. While fluid structure and dynamics under nano-confinement has been studied widely, studies addressing the effects of pore connectivity on fluid behavior under confinement are mostly limited to experimental [15, 16] or theoretical [17, 18] studies that lack molecular details.

Molecular simulations can provide important insights on these effects at the molecular level. For example, we have recently reported a systematic study on the effects of pore connectivity on the sorption of fluids using grand canonical Monte Carlo (GCMC) simulations [19]. While pore connectivity varied between 48 and 0 pore connections, was found to affect the sorption amounts, the effect of pore connectivity is expected to have a stronger effect on the dynamical properties [19]. In an earlier attempt, we used molecular dynamics (MD) simulations to address the effects of inter-connectivity of pores on the structure and dynamics of confined fluids by comparing the behavior of $CO_2$ and ethane in ZSM-22, a zeolite characterized by unidimensional pores of 0.5 nm diameter, with that in silicalite (all silica analogue of ZSM-5), a zeolite with pores of the similar size but characterized by unidimensional pores connected to each other via zig-zag channel like pores running in a perpendicular plane [14]. Connecting the one-dimensional pores with quasi one-dimensional channels was found to suppress the dynamics of both fluids while the effect of connecting the pores by artificially inserting two-dimensional inter-crystalline space was different for the two fluids [14]. The complex effects that connecting the pores may have on the dynamics of confined fluids may be difficult to capture in a simple comparative study employing two substrates with presence or absence of pore connectivity and therefore requires a systematic study probing a diverse range of heterogeneity in pore connectivity in the same substrate.

Silicalite is an all-silica analogue of ZSM-5 zeolite and has a network of one-dimensional channel-like pores inter-connected at regular intervals with quasi one-dimensional zigzag channels in a perpendicular plane [19]. This intricate pore network structure provides an opportunity to systematically vary the connections between pores by selectively blocking some of them, thus resulting in different scenarios of pore connectivity in the same substrate. Further, with different channel geometries – straight ellipsoidal channels oriented along the crystallographic *b*-axis and sinusoidal (zigzag) channels running in a parallel plane, selectively blocking all channels of one type can also help constrain the effects of pore geometry and tortuosity. In this case, the effect of tortuosity is limited to an on-off scenario (straight channels exhibit no tortuosity while zigzag channels have a tortuous shape). It is also worth noting that because the pore size in silicalite is similar regardless of pore type (straight versus zigzag), the pore size distribution is unimodal and sharply peaked. In the previous study addressing sorption properties [19], we created 12 models of silicalite with different degrees of pore connectivity by selectively blocking some channels via loading immobile methane molecules. These 12 models had pores connected via different combinations of connections ranging between 0 and 48.

In the present work we utilize these 12 silicalite models to study the effects of pore connectivity on the structure and dynamics of $CO_2$ and ethane confined within them using MD simulations. We begin by providing details of these models and the MD simulations in section 2, after which we present results from these studies in section 3. In section 4, we discuss the implications of the results of this study and in section 5 we present the salient conclusions.

## 2.0 SIMULATION DETAILS

### 2.1 Substrate Models

This study employed models of silicalite with different number of pore connections generated in a previous study [19]. In brief these models were made by selectively blocking some channels of a supercell of silicalite [21] made up of 2×2×3 unit cells obtained with VESTA [22]. Methane molecules were used as blockers which were treated as an immobile part of the silicalite substrate in the MD simulations described here. Figure 1 (a) illustrates the salient features of the pore network in silicalite. Straight channel-like pores are shown as pink cylinders while the zigzag channels connecting them are shown in blue. When some straight and/or zigzag channels are loaded with immobile methane molecules, they can isolate channels connected by them. This is illustrated in Figure 1 (b). Straight channels S3 and S4 are connected to each other by the zigzag channel Z4 (Figure 1(a)). When the zigzag channel Z4 is loaded with immobile methane molecules, it disrupts the connection between S3 and S4, isolating them from each other (Figure 1(b)). In the supercell of silicalite made up of 2×2×3 unit cells, there are a total of 48 connection combinations between straight and zigzag channels [19]. By selectively blocking some straight or zigzag channels, this number of pore connections can be reduced. The objective of this study is to investigate the effects of pore connectivity on the dynamics of fluids confined in a porous matrix. For this the system of pore connections were chosen so as to (i) vary the degree of pore connectivity over several values, and (ii) switch on/off pore tortuosity by selectively blocking all straight/zigzag channels respectively. This is illustrated in Figure 1 (c) which uses a cartoon schematic to represent the straight and zigzag channels by vertical magenta and horizontal blue lines, respectively. The number of intersections of these lines are proportional to the number of pore connections. A channel loaded with immobile methane is represented by an absence of the corresponding line. Thus, S4Z4 represents the unmodified silicalite where all straight and zigzag channels are open/free. When a fourth of all zigzag channels are blocked with immobile methane, it results in S4Z3 with the number of pore connections reduced accordingly. In general, a model substrate with n open straight channels and m open zigzag channels is represented by SnZm, where n and m represent the fraction out of a total of 4. All these substrates are illustrated in Figure 1 (c) and their pore-network details are listed in Table 1. For convenience, we classify the 12 substrates into 3 classes – S-major (substrates where more straight channels are open compared to zigzag channels; these are shown with red text color in the table), Z-major (substrates where more zigzag channels are open compared to straight channels, shown with blue text color in the table) and half-volume (substrate where half-the total pore volume available is blocked with methane, region in table 1 highlighted with yellow background). We note that the unmodified silicalite (S4Z4) as also some other models may belong to more than one of these 3 classes.

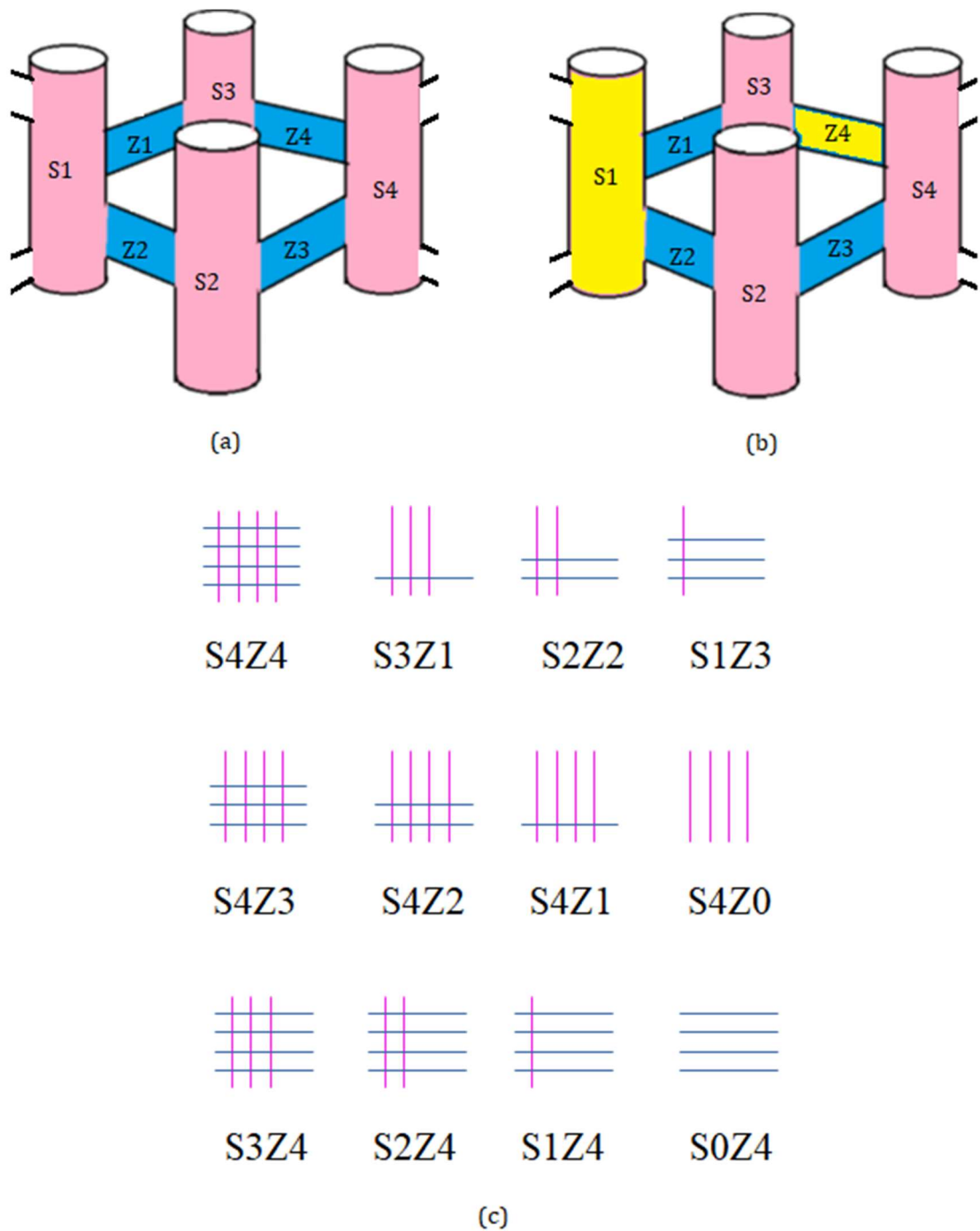

Figure 1. Cartoon schematics depicting (a) pore-network in unmodified silicalite. Straight channels (S1-S4) running in the crystallographic *b*-axis are shown as pink vertical cylinders. These are connected to each other via zigzag channels (Z1-Z4) shown in blue. Panel (b) shows the case when the connection between S1 and S4 is disrupted by loading Z4 with methane (yellow). All 12 model silicalite substrates used in the simulations are shown schematically in panel (c) representing the straight and zigzag channels by vertical magenta and horizontal blue lines respectively. The number of intersections of these lines are proportional to the number of pore connections. A channel loaded with immobile methane is represented by an absence of the corresponding line.

**Table 1.** Different systems simulated and the corresponding number of pore – interconnections. S4Z4 is the unmodified silicalite. Systems with a larger number of straight channels that are free compared to free zigzag channels are highlighted with red-colored text (S-major) while those with a larger number of free zigzag channels compared to free straight channels are highlighted in blue-colored text (Z-major). Systems where a half of the free space available in the unmodified silicalite is blocked/free are highlighted with yellow background and the system S2Z2 with half each of the straight and zigzag channels blocked/free is shown in green text color.

| System name | Open straight channels (% of total) | Open sinusoidal channels (% of total) | Number of pore connections | Number of $CO_2$ molecules | Number of ethane molecules |
|---|---|---|---|---|---|
| **S4Z4** | **100** | **100** | **48** | **128** | **128** |
| S4Z3 | 100 | 75 | 36 | 118 | 118 |
| S4Z2 | 100 | 50 | 24 | 110 | 102 |
| S4Z1 | 100 | 25 | 12 | 100 | 92 |
| S4Z0 | 100 | 0 | 0 | 92 | 78 |
| S3Z1 | 75 | 25 | 9 | 78 | 72 |
| S2Z2 | 50 | 50 | 12 | 76 | 70 |
| S1Z3 | 25 | 75 | 9 | 74 | 70 |
| S0Z4 | 0 | 100 | 0 | 78 | 75 |
| S1Z4 | 25 | 100 | 12 | 90 | 88 |
| S2Z4 | 50 | 100 | 24 | 104 | 102 |
| S3Z4 | 75 | 100 | 36 | 118 | 112 |

### 2.2. MD simulations

$CO_2$ or ethane molecules were loaded in the model silicalite at 308.16 K and 1 bar partial gas pressures using grand canonical Monte Carlo simulations carried out using DL-Monte [23]. These simulations provided the appropriate number of fluid molecules adsorbed in the supercell at the specified environmental conditions. The number of fluid molecules in the supercell are also tabulated in Table 1. Note that this approach of loading molecules is consistent for a given environmental condition of temperature and partial gas pressure and results in a different number of guest molecules between the two fluids in the same type of substrate. This difference is however, less than 18% in all cases, and is unlikely to result in significant deviation in properties due to loading differences. A larger difference in the number of adsorbed molecules can be found across different substrates for the same fluid. This is required to avoid spurious crowding due to a reduced pore volume in some substrates that would result if the same number of fluid molecules were used.

The configuration files obtained at the end of GCMC simulations were used as the starting configurations for MD simulations. As reported in an earlier publication, simulations of adsorption of $CO_2$ and ethane were carried out at 308.16 K because both guest fluids become supercritical at the highest partial pressure (100 bar) studied at this temperature. In continuation, the simulations in the present study were also carried out at the temperature of 308.16 K. Each MD simulation was carried out at in NVT ensemble using DL-Poly-4.10 [24]. As in the previous GCMC simulations [19], TraPPE-UA [25, 26] force field was used to model the interactions of $CO_2$, ethane, and the immobile methane molecules, while ClayFF [27] was used to represent silicalite atoms. The force-field parameters for all atoms involved are listed in Table 2. Cross-terms were calculated using the Lorentz-Berthelot mixing rules [28]. Using the same force-field for both fluids helps keep the formalism uniform. Further, as shown in a previous study [29], the set of force-fields used here (TraPPE-UA for the fluids and ClayFF for silicalite) reproduce the experimental adsorption isotherms of both fluids in silicalite. $CO_2$ and ethane were modeled as rigid molecules with translational and rotational degrees of freedom, while all atoms in the substrate including the blocker methane molecules were kept rigid throughout the simulation. For small guest molecules, the effects of using a flexible substrate are small enough to be ignored. For example, Newsome and Coppens [30] have studied $CO_2$ diffusion in Na-ZSM-5 an extra-framework cation containing analogue of silicalite using a rigid model of the substrate. They obtained diffusion coefficients of $CO_2$ in Na-ZSM-5 at 200 K and 300 K that were in fair agreement with neutron scattering experiments [31]. Further, the emphasis of this study is more on the variation of diffusion coefficient with the degree of pore connectivity instead of the absolute values of the diffusion coefficients. The gains in simplicity of the rigid framework used in this work thus outweigh a minor loss in accuracy. All simulations employed a calculation time step of 1 fs and lasted for 2 ns each, out of which the first 0.5 ns were used for equilibration. 0.5 ns period was found to be long enough for the temperature and energy to stabilize and the corresponding fluctuations reduce to acceptable values. The remaining 1.5 ns trajectory was used to calculate quantities with positions and velocities recorded after every 0.02 ps. For maintaining the system temperature in the NVT simulations, Nosé-Hover thermostat with a relaxation time of 1 ps was used.

Table 2: Force-field parameters used in the simulations. $O_c$ stands for the oxygen atom belonging to $CO_2$ molecule.

| Species | Atom/Psuedo-atom | Charge (q/e) | ε (kJ) | σ (Å) |
|---|---|---|---|---|
| Silicalite | Si | +2.10 | 0.000008 | 3.301 |
|  | O | -1.05 | 0.650198 | 3.166 |
| $CO_2$ | C | +0.70 | 0.224 | 2.80 |
|  | $O_c$ | -0.35 | 0.657 | 3.05 |
| Ethane | $CH_3$ | 0.00 | 0.815 | 3.75 |
| Blocker Methane | $CH_4$ | 0.00 | 1.231 | 3.73 |

**3.0 RESULTS**

We have already discussed the sorption and structural properties of $CO_2$ and ethane in these models in a previous publication [19]. Here we focus exclusively on the dynamical properties of the confined fluids, specifically translational, rotational and the intermolecular modes of vibration. We note that the intramolecular modes of vibration in both fluids are frozen by design due to a selection of rigid models for both fluids.

**3.1** Translational motion

The translational motion of the confined fluids is described in terms of the mean squared displacement of the center of mass of the molecules. Figure 2 shows the MSD vs time plots for both fluids in all silicalite models investigated. In general, ethane mobility is found to be faster compared to $CO_2$ and connecting the pores is found to enhance the diffusivity of both fluids. For both fluids, a combination of straight and zigzag channels is found to facilitate diffusion. For $CO_2$ straight channels allow faster diffusion compared to zigzag channels, whereas for ethane, the distinction between the two channel types in facilitating diffusion is less clear. When the pore volume in silicalite is reduced to half, an equal distribution of the pore volume between straight and zigzag channels (S2Z2) is found to best facilitate the diffusion of both fluids.

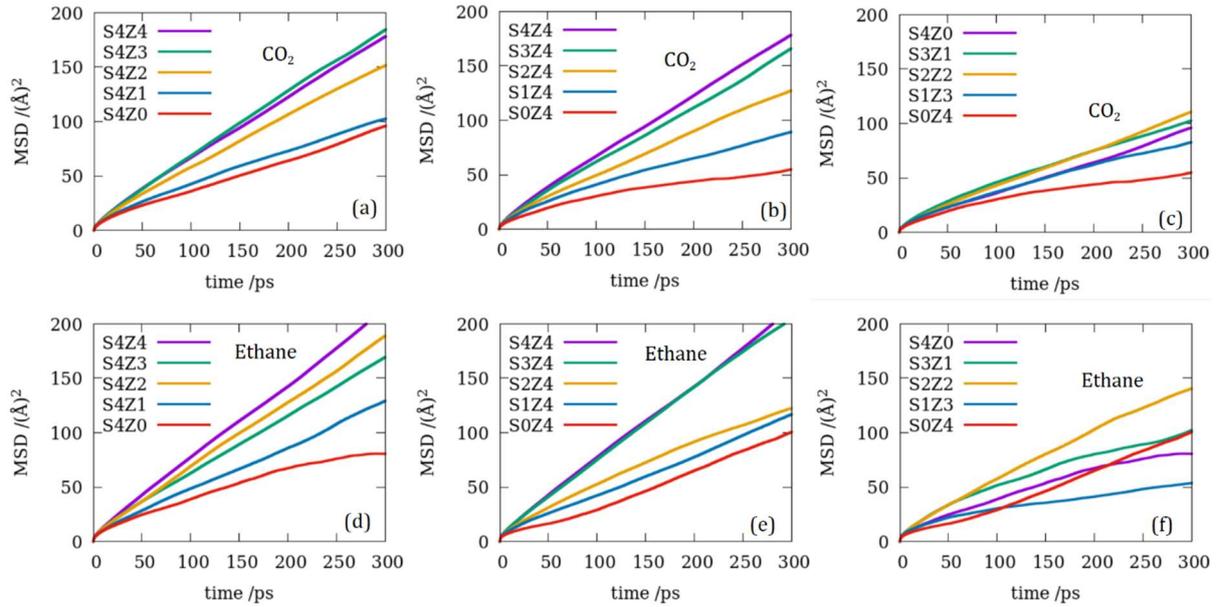

**Figure 2.** Mean squared displacement (MSD) as a function of time for $CO_2$ (a-c) and ethane (d-f) in silicalite. Data in (a) and (d) show S-major models of silicalite, where the number of open straight channels is equal to or greater than those of open zigzag channels. Panels (b) and (e) show the data for Z-major models where the number of zigzag channels is larger or equal to that of straight channels and (c) and (f) show the case of models with half of the unmodified silicalite (S4Z4) blocked with methane molecules. For meaning of the nomenclature SnZm refer to the text and Table 1.

Diffusivity of fluids can be quantified by calculating the self-diffusion coefficient ($D_{self}$) from MD simulation data using the Einstein relation [28],

$$D_{self} = \frac{1}{2n_d}\left(\lim_{t\to\infty}\frac{MSD}{t}\right) \qquad (1),$$

where $n_d$ is the number of degrees of freedom of translational motion and the quantity in the parentheses is the slope of MSD vs time plots evaluated at long times. We calculated the diffusion coefficient from the slope of MSD vs time plots in three different time ranges of 100 – 200 ps, 150 – 250 ps and 200 – 300 ps. The values of the diffusion coefficients obtained in three different time ranges thus were averaged and uncertainty obtained as the standard deviation over these averages. Values of $D_{self}$ calculated using Eq. 1 and the data in Figure 2 are shown in Figure 3 as a function of the number of pore connections. In general, the effect of connecting the pores is to enhance the diffusivity of both fluids. Further, in the case of $CO_2$, diffusivity in the S-major substrates is higher than that in the Z-major substrate, a consequence of the simpler geometry of the straight channels. The variation in the diffusivity of ethane in different model substrate is relatively more irregular. Ethane diffusion through zigzag channels is comparable with that in the straight channels and this results in an overall higher diffusivity of ethane in the unmodified silicalite (S4Z4) also observed in an earlier study [32].

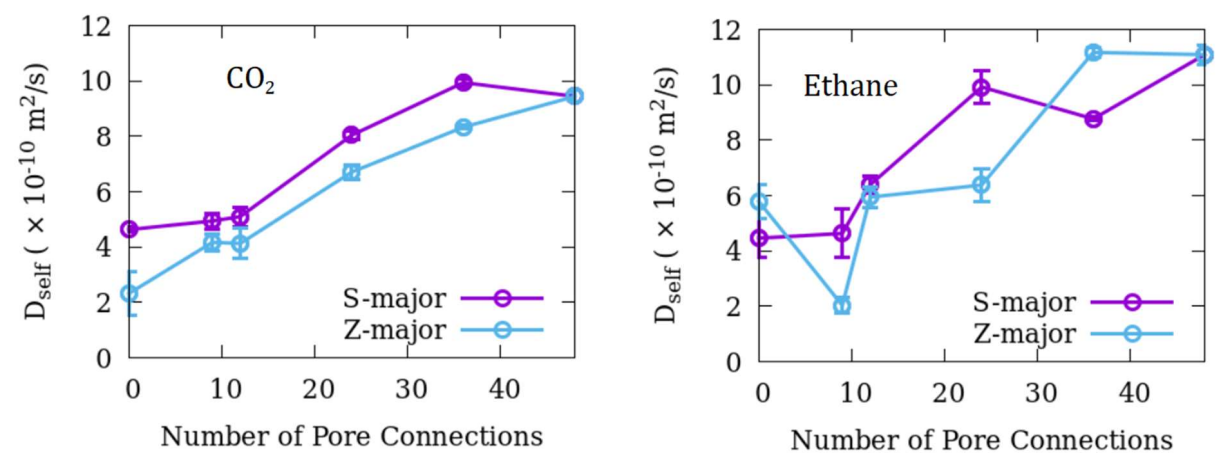

**Figure 3.** Self-diffusion coefficients ($D_{self}$) as a function of the number of pore connections for $CO_2$ and ethane confined in silicalite.

Because of the intricate pore network in silicalite, the dynamics of confined fluid is expected to be anisotropic [32]. This results in a separation of the MSD in different Cartesian directions. As straight channels are oriented exactly along the Y-direction, whereas the zigzag channels lie in the X-Z plane oriented roughly, but not exactly along the X-axis, the MSD of fluids in silicalite follows the order MSDy > MSDx > MSDz. In case of isolated straight (S4Z0) or zigzag (S0Z4) channels, the pore network is oriented completely in the Y direction and the X-Z plane, respectively, and the MSD components along different directions can show significant variation. In Figure 4, we show the overall MSD resolved along the three Cartesian directions for the three model substrates S4Z4, S4Z0 and S0Z4. As expected, the MSD along Y direction in S4Z4 is highest for both fluids followed by that along the X direction. In S4Z0, the motion occurs exclusively along the Y direction and the MSD in the X and Z-directions is negligible. Conversely, in S0Z4, the motion is predominantly along X-direction.

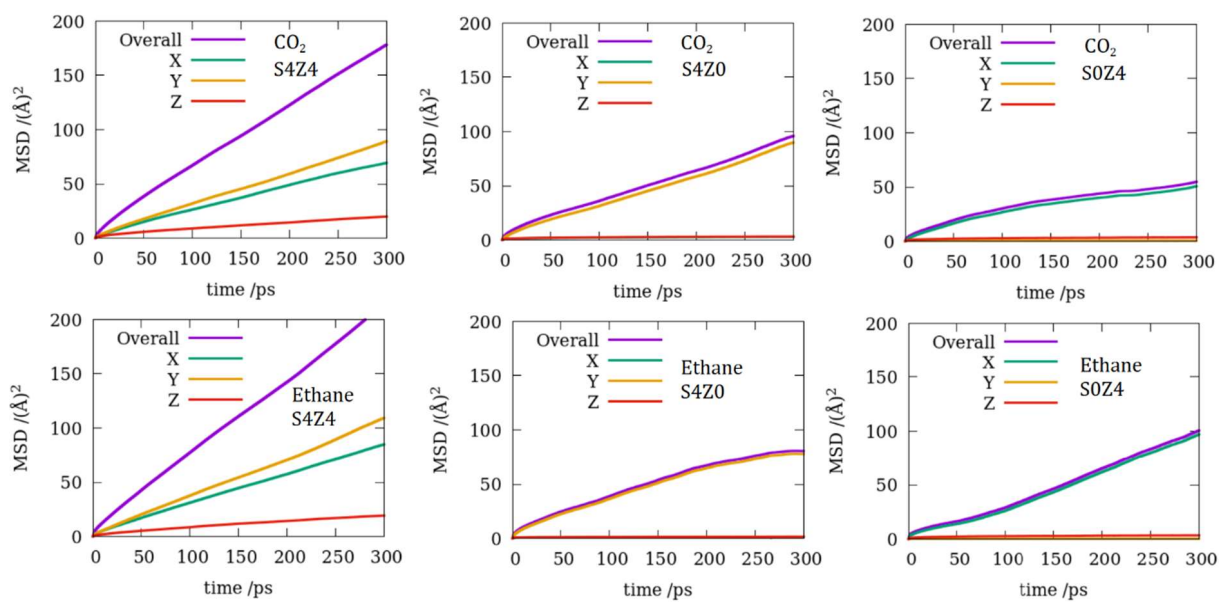

**Figure 4**. MSD resolved in Cartesian directions compared to the overall MSD. Upper panels show $CO_2$ while lower panels show ethane data. For clarity, data are shown for only three substrates – S4Z4, S4Z0 and S0Z4.

Figures 5 -7 show the direction specific self-diffusion coefficients of $CO_2$ and ethane in silicalite as a function of pore connectivity. These direction specific self-diffusion coefficients were calculated using long-time slope of the corresponding MSD vs time plots in Eq. 1 with $n_d$=1. For $CO_2$, the diffusion coefficients in X and Z direction increases with increasing number of pore connections for both S-major as well as Z-major substrates. In the Y-direction, $CO_2$ diffusivity is clearly facilitated by pore connectivity in Z-major substrates, whereas for S-major substrates, this effect is less clear with greater variation of $D_{self}$ with number of pore connections. The difference between the two extreme cases of S4Z0 (leftmost datum) and S4Z4 (rightmost datum) is smaller than the uncertainties involved. This similarity in the diffusivity along Y-axis in S4Z0 and S4Z4 is also true for ethane, where the difference between the data representing the two substrates is negligible. For all other cases with ethane, the pore connections can be seen to facilitate diffusion, except for diffusion along X-direction in Z-major substrates where the variation is complex and the difference between the extreme cases is negligible.

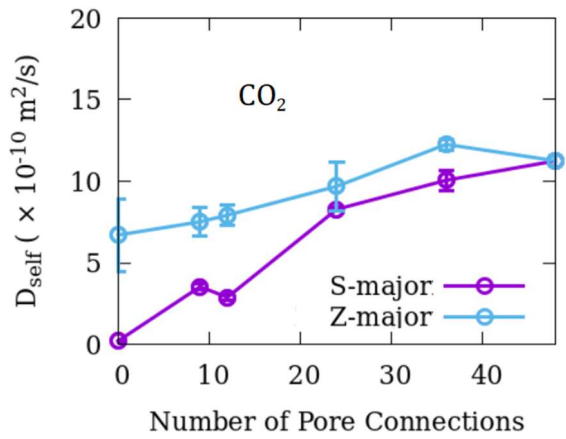
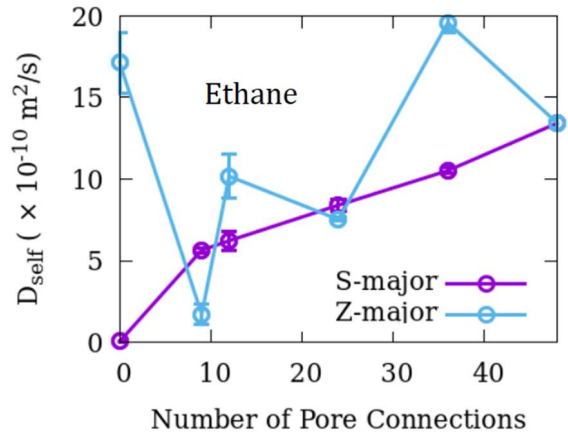

**Figure 5**. Direction-specific self-diffusion coefficients along Cartesian X-direction of $CO_2$ and ethane in silicalite as a function of pore connectivity.

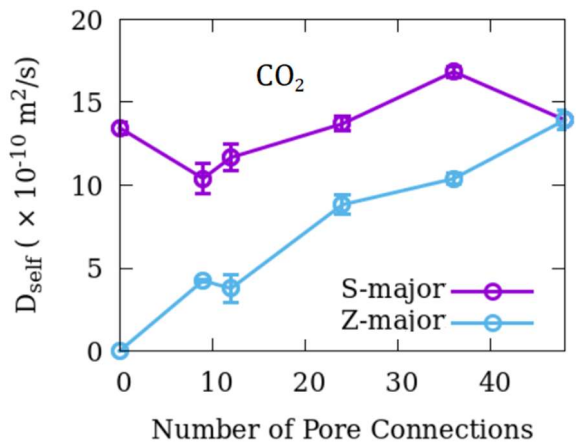
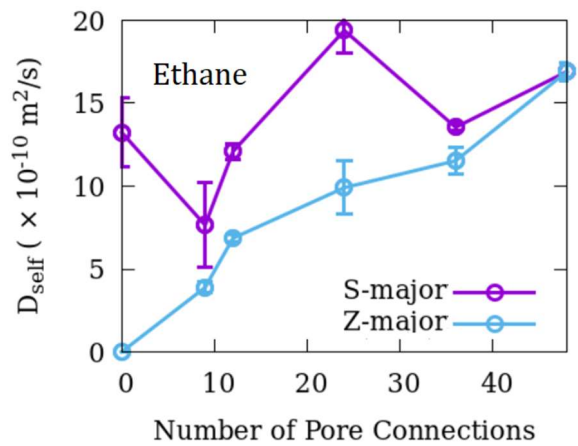

**Figure 6**. Direction-specific self-diffusion coefficients along Cartesian Y-direction of $CO_2$ and ethane in silicalite as a function of pore connectivity.

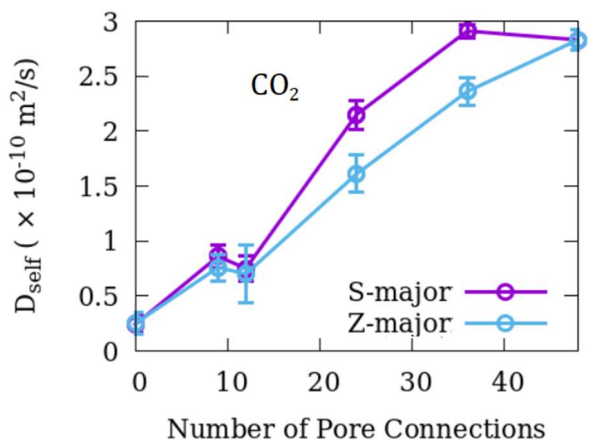
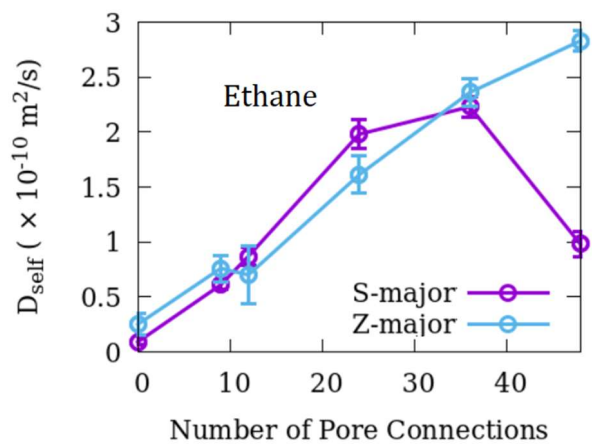

**Figure 7**. Direction specific self-diffusion coefficients along Cartesian Z-direction of $CO_2$ and ethane in silicalite as a function of pore connectivity.

In the TraPPE-UA formalism, both $CO_2$ and ethane are linear dumbbell shaped molecules. For such molecules, one can expect a difference in their dynamical behavior along the molecular axis and in a plane perpendicular to it [33]. As the molecules rotate, their orientation changes in time. At each time frame, we resolve the displacement of the molecule along directions normal (N) and parallel (P) to the instantaneous molecular axis. Since two directions can be normal to the molecular axis, and only one is parallel to it, the MSD perpendicular to the molecular axis is normalized via a division by 2. Figure 8 shows the MSD in the two directions with respect to the molecular axis for $CO_2$ and ethane in 3 model silicalite substrates. While $CO_2$ shows a little preference in the direction of motion with respect to molecular axis, ethane clearly prefers to move parallel to the molecular axis in all the three cases investigated.

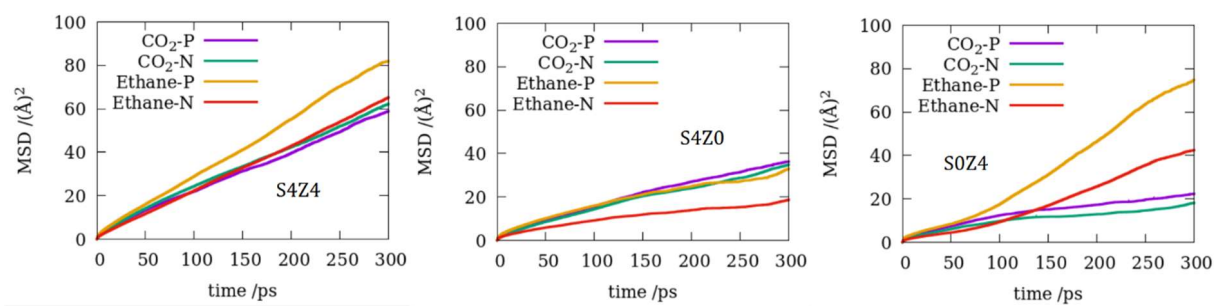

**Figure 8**. MSD along the molecular axis (P) and normal to it (N) for the two fluids in three model silicalite substrates.

**3.2** Rotational Motion

Rotational motion of the two fluids in silicalite is studied via rotational correlation function (RCF) calculated using Eq. 2 [32]

RCF=<$u(t+t_0)$. $u(t_0)$>    (2)

Here $u(t)$ is a unit vector attached to the molecular axis at time $t$ and the angular brackets denote an average over all molecules and the times of origin $t_0$. RCF of $CO_2$ and ethane in the silicalite substrates are shown in Figure 9. As in translational motion, ethane is found to exhibit a faster rotational motion in silicalite. This is evident from a faster decay of RCF of ethane. Further, the ethane RCF exhibits a strong wiggle at around 1 ps, - a distinct signature of restricted librational motion at short times [32, 34]. Over a longer timescale of a few tens of picoseconds the RCF decay completely. The RCF behavior in both fluids undergo a transition at around 1 ps. Below this time scale, the decay is fast and independent of the substrate. This is the initial free rotation that occurs before the molecule can experience any hindrance to rotation due to its environment – e.g., substrate atoms, other fluid molecules, etc. – and is characteristic of the fluid molecular properties alone.

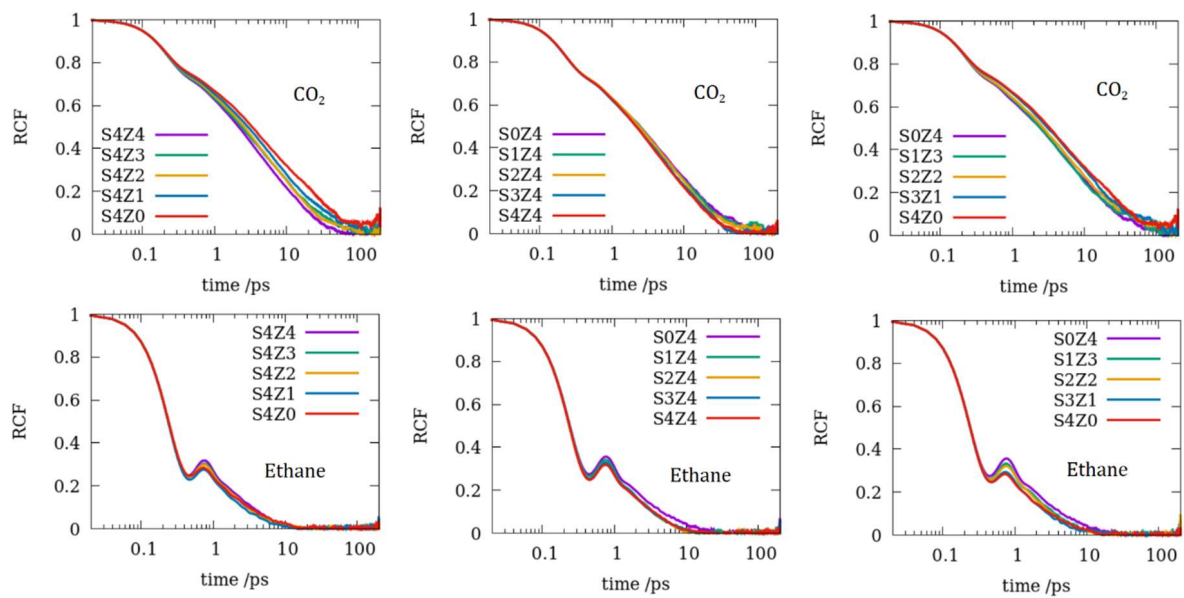

**Figure 9**. Rotational correlation functions (RCF) of $CO_2$ (top panels) and ethane (bottom panels) in various models of silicalite substrate.

Beyond 1 ps, the RCF of both fluids decay at a relatively slower rate and also show a dependence on the substrate. This region can therefore be used to study the effects of pore connectivity and tortuosity on the rotational behavior of the two fluids. In particular, the RCF of both fluids were fit with the following exponential decay function to obtain a characteristic decay time ($\tau$)

$$RCF = a*\exp(-t/\tau) + b \quad (3)$$

The RCF were fitted with the above model function with a, b and $\tau$ as the fitting parameters. In Figure 10, we show the values of the decay time $\tau$ obtained for all substrates as a function of pore connectivity and tortuosity (in terms of the percentage of open straight channels). Like translational motion, the rotational motion of both fluids is found to be facilitated by connecting the pores, although the effect of pore connectivity is relatively smaller on rotational motion as compared to translational motion. By comparison, rotational motion of $CO_2$ is affected by pore tortuosity somewhat more than that of ethane. Further, while $CO_2$ rotation gets faster (smaller times for zigzag channels represented by the leftmost datum, 0 straight channels) in tortuous pores, the rotation of ethane is slightly suppressed in them compared to straight channels.

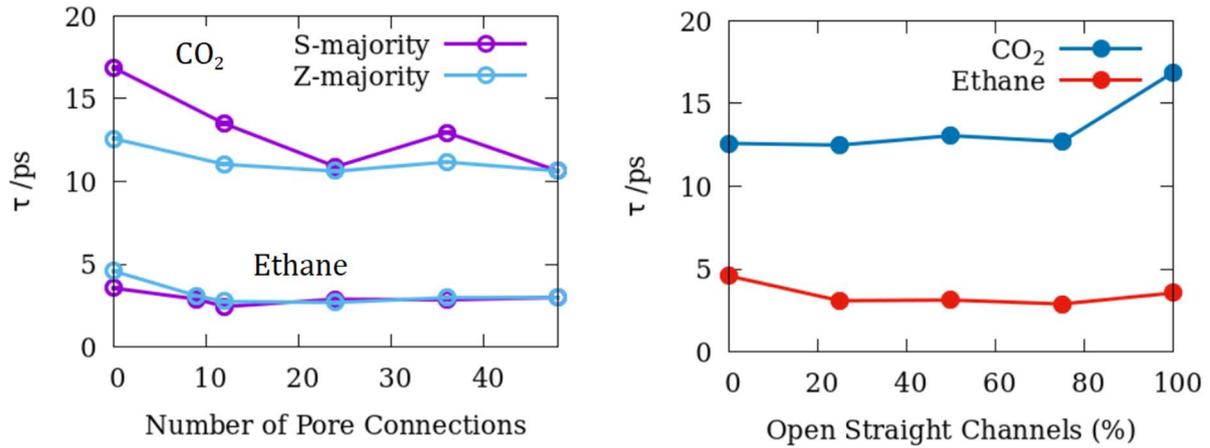

**Figure 10**. Characteristic decay times τ for rotational motion obtained by fitting the RCF with an exponential decay function (Eq. 4), for $CO_2$ and ethane in silicalite as a function of pore connectivity and the percentage of open straight channels. Uncertainty in the data is smaller than the symbols used.

### 3.3 Intermolecular Vibration

Intermolecular vibrational motion is studied via the power spectra I(ω) of the velocity autocorrelation function (VACF) using the following relations [32, 35]

VACF=<$v(t+t_0)$. $v(t_0)$>     (4)

I(ω) = ∫(VACF)cos(ωt)dt     (5)

The effect of pore connectivity on the intermolecular vibrational spectra of $CO_2$ in silicalite is negligible (Figure 11). However, comparing the half-volume data (rightmost panel), the peak in the $CO_2$ spectra shifts slightly to higher energies when more zigzag channels are open. For ethane, the power spectra are bimodal for all substrates and the higher energy mode increases at the expense of the lower energy mode as more zigzag channels are opened. Thus, the intermolecular vibrations of both fluids are more energetic in zigzag channels as compared to straight channels.

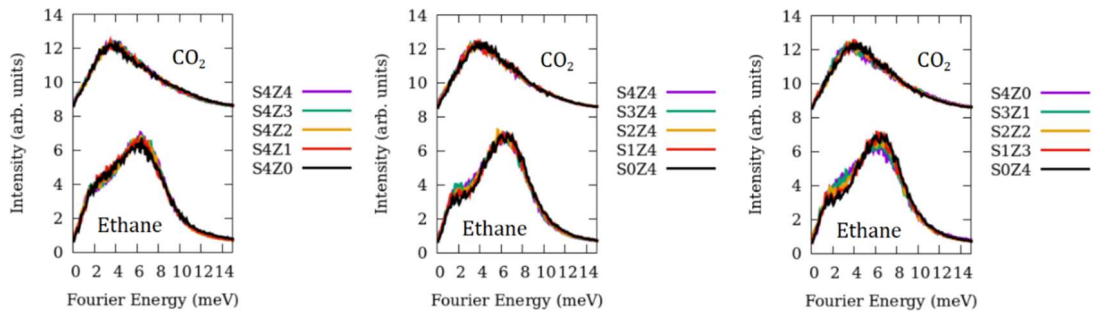

**Figure 11**. Power spectra corresponding to the intermolecular vibrational motion of $CO_2$ (top) and ethane (bottom) in different silicalite substrates. For clarity, the intensities of the $CO_2$ spectra are shifted upwards by 8 units.

## 4.0 DISCUSSION

In what follows we discuss the effects of pore connectivity and tortuosity on the dynamical behavior of two fluids and their implications.

### 4.1 Effects of Pore Connectivity

The overall effect of connectivity in the pore network of silicalite is to enhance both the translational as well as the rotational motion of both fluids – $CO_2$ and ethane (Figures 3 and 10 (a)). The effect of connecting straight or tortuous zigzag pores on the dynamics of fluids through them is however more nuanced. As these pores are essentially one-dimensional channels, a fair comparison between the effects of connecting them can be made by comparing the 1-dimensional diffusivity of the confined fluids along the direction of the channel axis. This is relatively easier to do for straight channels as they are perfectly aligned along the Cartesian Y-direction whereas zigzag channels have their pore axis aligned mostly along the X-direction with a minor component along the Z-direction. Figure 6 presents diffusivity along Y-direction, where the data corresponding to S-major substrates exhibit an irregular behavior while that corresponding to Z-major substrates exhibit a relatively smooth increase with an increase in pore connections. This, along with the results shown in Figure 3 indicates that while connecting the straight channels does increase the overall diffusivity, this increase is a result of the cross-current made available by opening a cross-channel while the motion along the original direction remains largely unaffected. Note that the difference between the diffusivity along Y-direction for totally isolated straight channels and straight channels connected to maximum is negligible for both fluids. It is also noteworthy that in a previous study [14] a comparison between the 1-dimensional diffusivity of both $CO_2$ and ethane in isolated straight channels of ZSM-22 and connected straight channels of ZSM-5 led us to conclude that connecting the pores suppresses the 1-dimensional translational diffusivity of both fluids. While the previous comparative study employed two different substrates with similar shape and size of channel-like pores, the present study employing systematic blocking of the connections between straight channels in the same substrate avoids contaminating the effects seen by factors other than pore connectivity *viz.* a slight difference in pore size and shape. Like the implications of Figure 6, Figure 5 shows that while connecting straight channels progressively leads to a clear enhancement in the diffusivity along X-direction (perpendicular to the straight channel axis) for both fluids, the enhancement achieved in the diffusivity along this direction on connecting the zigzag channels is less pronounced for $CO_2$. For ethane, the variation of diffusivity along X-direction in Z-major substrates is irregular with no significant difference between the two extreme cases of isolated zigzag channels and those connected to full extent.

Examination of the variation of overall and direction-specific diffusion coefficients reveals that ethane exhibits a more complex pattern whereas data corresponding to $CO_2$ display a relatively smoother variation. In Figure 12 we show trajectories of the center of mass of one molecule each of $CO_2$ and ethane over the entire production time of 1.5 ns. Note that the shape traced by the trajectory of the $CO_2$ molecule is wider than that of ethane. This implies that $CO_2$ molecules are distributed more widely across the pore space and are in general closer to the pore surface than ethane molecules. This was also seen in the distribution observed in the previous GCMC simulations and is a result of a stronger substrate-fluid interaction in the case of $CO_2$. This leads to an overall smaller diffusivity in $CO_2$ compared to ethane. Further, at the pore intersections (blue encircled region in the left-most panel), $CO_2$ molecules can occupy a wider region facilitating a smooth inter-channel migration whereas for ethane the inter-channel migration takes place via a relatively narrower intersection. This restricted inter-channel mobility might be responsible for the greater variation in the diffusivity of ethane as the pore connections are varied.

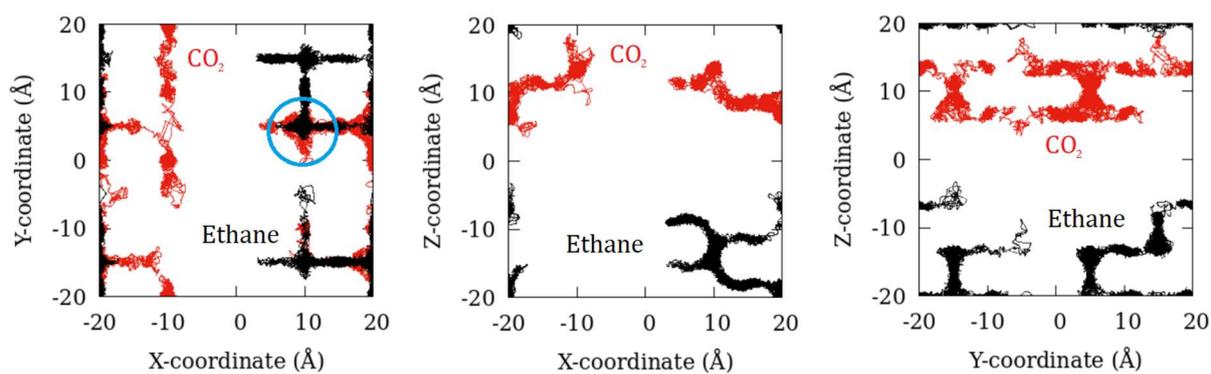

**Figure 12**. Trajectory of the center of mass of a tagged $CO_2$ (red) and ethane (black) molecule over 1.5 ns in S4Z4. A pore intersection is highlighted in the left-most panel by encircling in blue.

Compared to translational motion, the effects of pore connectivity on the rotational motion are only marginal and are observed to be slightly stronger on rotation of $CO_2$ in the S-major substrates as compared to Z-major substrates (Figure 10). This is because, compared to translational motion, the rotational motion occurs at a much smaller length scale. Figure 13 shows the rotational motion of a tagged $CO_2$ and ethane molecule each in the center of mass frame over a period of 10 ps. Over this small interval of time, while the ethane molecule has traversed the entire orientational space available (yellow sphere) the $CO_2$ molecule is restricted to only a limited part of the orientational space available to it (cyan sphere). This is consistent with the fact that the RCF for ethane decay completely within 10 ps while that of $CO_2$ have non-zero correlations left at this time (Figure 9) and indicates a stronger restriction on the rotational motion of $CO_2$ compared to ethane leading to smaller rotational time scales for the latter (Figure 10). This is a consequence of stronger substrate fluid interactions for $CO_2$ compared to ethane noted earlier. The tortuous shape of the zigzag channel can facilitate the rotational motion of $CO_2$ to a larger extent as these molecules lie relatively closer to the pore surface and therefore are affected by the pore shape to a greater extent. This results in a faster rotation as more and more zigzag channels are opened in S-major substrate leading to greater pore connectivity.

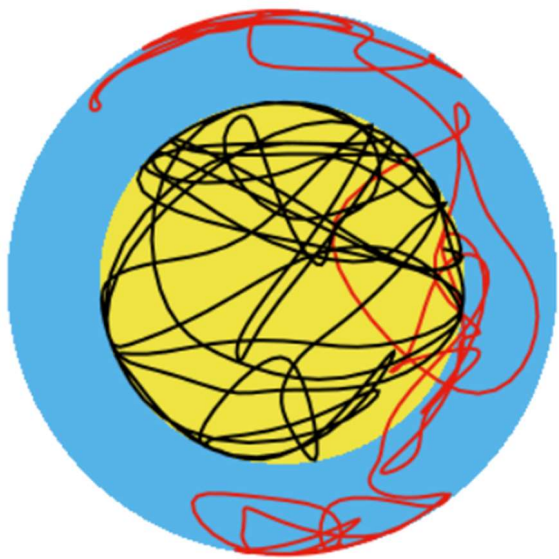

**Figure 13**. Trajectory of an oxygen atom belonging to a tagged $CO_2$ molecule (red) and that of a $CH_3$ psuedo atom belonging to a tagged ethane molecule (black) each in the center of mass frame of reference over a time of 10 ps in S4Z4. The entire orientational space available for ethane is shown as yellow sphere of diameter 0.154 nm while that for $CO_2$ is shown as the cyan sphere of diameter 0.232 nm containing the yellow sphere within itself.

4.2 Effects of Pore Tortuosity

The effects of pore tortuosity can be studied by considering the variation of properties in the substrates characterized by 'half-volume' (yellow highlighted region in Table 1). This is shown in Figure 10 for rotational time scales, Figure 11 for intermolecular vibrational modes, and Figure 14 for self-diffusion coefficients corresponding to the translational motion. While the overall translational motion of $CO_2$ is clearly faster in straight channels and gets negatively impacted by tortuosity of zigzag channels, for ethane, no significant difference is observed between the diffusivity in different channel types. This is also a consequence of the stronger fluid-substrate interaction for $CO_2$ which lies closer to the pore surface and is therefore impacted by the pore shape to a greater extent. For the same reason, the rotational motion of $CO_2$ is faster in the zigzag channels (Figure 10, right panel) as the tortuous shape of these channels facilitate the rotational motion as mentioned in the previous section. The tortuous shape of the zigzag channels also enhances the energy of the intermolecular vibrations (Figure 11).

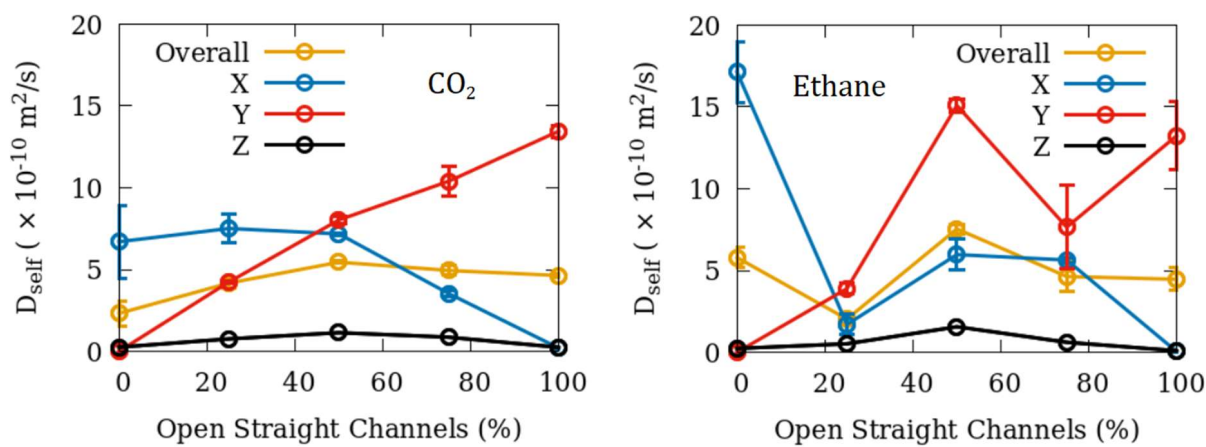

**Figure 14**. Overall and direction-specific self-diffusion coefficients corresponding to translational motion of $CO_2$ (left) and ethane (right) in silicalite substrates characterized by 'half-volume' (yellow highlighted region in Table 1). The left-most data points in both panels represent the highest number of tortuous channels while the right-most data points represent the case of straight channels with no tortuosity.

4.3 Implications

The results presented here can have important implications for applications that use fluids under confinement. While connecting several pores can enhance the overall diffusivity of the confined fluids, a judicious selection of different pore shapes in the same material can give rise to interesting effects that can be beneficial. For example, silicalite with a set of straight channels in one direction and a set of channels with a tortuous shape in the perpendicular direction can be used to separate different attributes by using well oriented samples. Further, the slowing of translational diffusivity of $CO_2$ in the zigzag channels can be used in engineered oriented silicalite membranes that let $CO_2$ pass through it more efficiently in one direction compared to a perpendicular plane.

In addition to the effects of pore-connectivity and tortuosity, the use of two representative fluids – one apolar and other quadrupolar also helps understand the role of fluid properties in these effects. While $CO_2$ with stronger interactions with the host due to a quadrupolar moment is affected to a greater extent by pore tortuosity, ethane remains less affected due to a weaker interaction. This suggests an important difference between apolar and polar molecules that can be utilized in separation processes. In a mixture of polar and apolar fluid, the polar fluid can be expected to be affected to a greater extent by pore tortuosity thereby enhancing the difference in the mobility of these fluids. Such a mixture can therefore be separated by making oriented membranes with tortuous pores aligned along the flow direction.

## 5.0 CONCLUSIONS

By selectively blocking some pore spaces of silicalite via immobile methane molecules, molecular dynamics simulations of silicalite substrates are used to systematically study the effects of pore connectivity and pore tortuosity on the dynamics of two representative carbon-bearing fluids – $CO_2$ and ethane. While overall diffusivity of both molecules is enhanced as more pores are connected, the trends observed in the 1-dimensional diffusivity along the channel direction are more nuanced. Rotational and intermolecular vibrational motions of both fluids are found to be impacted on connecting the pores to a relatively smaller extent compared to the translational motion. Pore tortuosity is found to facilitate the rotational motion of $CO_2$ and suppress its translational motion, while its effects on the motion of ethane are less pronounced. The intermolecular vibrations of both fluids are found to be more energetic in the tortuous zigzag channels. The results of this study can be used to design oriented silicalite membranes that are more efficient in permitting $CO_2$ molecules to pass in one direction compared to a perpendicular direction. Further membranes with tortuous pores aligned along the direction of flow can be used to separate a mixture of polar and apolar fluids based on the difference in their mobilities.

## 6.0 AUTHOR CONTRIBUTIONS

S. G. Conceptualized the study, carried out the simulations, analyzed and interpreted the simulation data, and wrote the original draft. D.C. secured the funding. Both authors reviewed and revised the initial manuscript draft

## 7.0 ACKNOWLEDGEMENT

This research was funded by the U.S. Department of Basic Energy, Office of Science, Office of Basic Energy Sciences, Division of Chemical Sciences, Geosciences and Biosciences, Geosciences Program, grant number DESC0006878. We would like to acknowledge STFC's Daresbury Laboratory for providing the package DL-Poly, which was used in this work. Simulations reported in this work were carried out at the College of Arts and Sciences (ASC) Unity Cluster of the Ohio State University. The computational resources and support provided is gratefully acknowledged (Sandy Shew, Brent Curtiss, Keith Stewart and John Heimaster). Figures in this manuscript were made using the freely available plotting software Gnuplot [36] (Figures 2–14).

## 8.0 CONFLICT OF INTEREST STATEMENT

The authors declare that there is no conflict of interest.

## 9.0 DATA AVAILABILITY

All data related to this study are available in this article.